\documentclass[fleqn,10pt]{wlscirep}
\usepackage[utf8]{inputenc}
\usepackage[T1]{fontenc}
\title{Unidirectional Zero Reflection and Perfect Absorption via Exceptional Points in Active Piezoelectric Willis Metamaterials}

\author[1]{Hrishikesh Danawe}
\author[1,*]{Serife Tol}
\affil[1]{Department of Mechanical Engineering, University of Michigan, Ann Arbor, MI USA 48109}
\affil[*]{Corresponding author(s): Email(s) stol@umich.edu}

\begin{abstract}
Electro-momentum coupling in piezoelectric metamaterials with broken inversion symmetry enables asymmetric elastic wave transport by linking macroscopic electric fields to momentum, an effect analogous to Willis coupling in elastic media. A one-dimensional layered piezoelectric metamaterial integrated with shunt circuits, consisting of a resistor, inductor, and strain-proportional voltage feedback gain, is proposed to achieve dynamic control of frequency-dependent stiffness and damping through electromechanical interactions. Tuning the circuit parameters yields direction-dependent wave scattering at targeted frequencies. Dynamic homogenization reveals macroscopic constitutive relations exhibiting both Willis and electro-momentum couplings. Non-Hermitian exceptional points are identified, where scattering eigenmodes coalesce and produce extreme asymmetries in wave response. Near these points, the system realizes unidirectional zero reflection (UZR) and unidirectional perfect absorption (UPA), achieving complete absorption from one direction and total reflection from the opposite side. The findings demonstrate a compact and reconfigurable platform for tunable, directional elastic wave control using passive-active hybrid metamaterials, opening new avenues for programmable devices in acoustic isolation, wave-based computing, sensing, and energy manipulation in solid media.
\end{abstract}
\begin{document}

\flushbottom
\maketitle
% * <john.hammersley@gmail.com> 2015-02-09T12:07:31.197Z:
%
%  Click the title above to edit the author information and abstract
%
\thispagestyle{empty}

\section*{Introduction}

Nonreciprocal wave propagation is a fascinating area of physics that challenges the conventional symmetry typically associated with wave transmission, implying that the system’s response remains invariant under source–receiver interchange. This reciprocity arises from the assumptions of time-reversal symmetry, linearity, and time-invariance of the medium~\cite{Auld,achenbach2003reciprocity,Maznev2013}. In contrast, nonreciprocal systems intentionally break this symmetry, allowing waves to propagate preferentially in one direction while being attenuated, reflected, or entirely suppressed in the opposite direction. Such direction-dependent control of wave transport offers transformative opportunities in the design of next-generation acoustic and elastic devices, including unidirectional sensors, isolators, signal routers, and energy-harvesting systems. Traditional strategies to achieve nonreciprocal wave transport often rely on mechanisms such as nonlinear interactions~\cite{Liang2009,Boechler2011,Yousefzadeh2021,Popa2014}, spatiotemporal modulation~\cite{Trainiti2016,Nassar2017,Nassar2020}, or moving media~\cite{Fleury2014}. While these approaches have enabled several groundbreaking demonstrations, their practical deployment is often hindered by complexity, requirements for dynamic biasing, or scalability constraints. An emerging alternative leverages engineered metamaterials, particularly those incorporating spatial asymmetry and tailored loss or gain, to realize asymmetric wave responses in linear, time-invariant systems.

Within this paradigm, Willis metamaterials~\cite{milton2007modifications,pernas2020fundamental,sieck2017origins}, characterized by microstructural asymmetry and bianisotropic constitutive relations, offer a powerful framework for asymmetric wave control. Despite supporting exotic couplings such as stress–velocity and momentum–strain, these materials remain fundamentally reciprocal due to the preservation of time-reversal symmetry and passivity~\cite{muhlestein2016reciprocity,cho2021acoustic}. Nevertheless, they can exhibit asymmetric reflection phases for waves incident from opposite directions, owing to direction-dependent impedance introduced by Willis coupling~\cite{liu2019willis,hao2022experimental}. While the transmission remains symmetric as required by reciprocity, the phase of the reflected waves can differ based on the incidence direction. However, to achieve strong asymmetry in the reflection amplitude, it is typically necessary to introduce loss, thereby rendering the system non-Hermitian~\cite{liu2019willis,hao2022experimental}. By carefully engineering loss, reflection in one direction can be strongly suppressed, approaching unidirectional zero reflection (UZR), while reflection in the opposite direction remains significant~\cite{merkel2018unidirectional,liu2019willis}. This phenomenon is closely linked to the concept of exceptional points (EPs) in non-Hermitian physics, where eigenvalues and eigenvectors of the scattering matrix coalesce, leading to abrupt transitions and extreme asymmetries. However, directly tuning loss in physical materials is often impractical.

To overcome this challenge, piezoelectric Willis metamaterials enable external circuit-based control of loss and resonance. In our previous work, we introduced a resistive-inductive (RL) shunt circuit to realize tunable asymmetric wave phenomena via externally tailored dissipation and resonance~\cite{danawe2023electro}. These systems leverage strong electromechanical coupling, which enables an additional bianisotropic term between electric field and momentum, termed electro-momentum coupling~\cite{pernas2020symmetry}, an electromechanical analog to Willis coupling. Both Willis and electro-momentum couplings contribute to directional wave behavior in piezoelectric systems~\cite{pernas2021electromomentum}. Such couplings can be significantly enhanced through topology optimization~\cite{huynh2023maximizing,lee2022maximum,zhang2022rational,huynh2025effect}. Combined with mechanical loss in the host structure, these systems can exhibit extreme reflection asymmetry and UZR~\cite{huynh2025effect}. Furthermore, the inclusion of tunable resistance enables frequency-selective UZR~\cite{wu2022versatile}. However, achieving both UZR and unidirectional perfect absorption (UPA), where all incoming energy from one direction is absorbed and reflection is entirely suppressed, remains an open challenge for passive elastic systems. To the best of our knowledge, there are no demonstrations of simultaneously achieving UZR and UPA in such systems via dissipation alone. Nonetheless, the electromechanical coupling in piezoelectric metamaterials enables further control through external feedback circuits, offering a pathway to wave control that surpasses the limitations of passive configurations.

In this work, we demonstrate that piezoelectric metamaterials composed of stacked layers integrated with tunable shunt circuits featuring strain-proportional voltage feedback, inductive resonance, and resistive damping that can realize non-Hermitian scattering conditions, including exceptional points that yield highly asymmetric wave responses. By tuning the external circuit parameters, we gain precise control over the system’s stiffness and loss characteristics, enabling directional scattering tailored to specific frequencies. Using dynamic homogenization, we extract the macroscopic effective properties of the one-dimensional layered metamaterial, revealing the presence and tunability of both Willis and electro-momentum couplings. This circuit-level programmability allows the system to exhibit unidirectional zero reflection (UZR) and unidirectional perfect absorption (UPA), along with retroreflection in the reverse direction. By leveraging strong electromechanical interactions and active–passive hybridization, our platform offers a compact, reconfigurable, and scalable solution for nonreciprocal elastic wave manipulation, opening new avenues for wave-based signal routing, isolation, and adaptive control in engineered solids.

\section*{Results}
\subsection*{Design of Willis metamaterial}
\begin{figure}[ht]
    \centering
    \includegraphics[width=\linewidth]{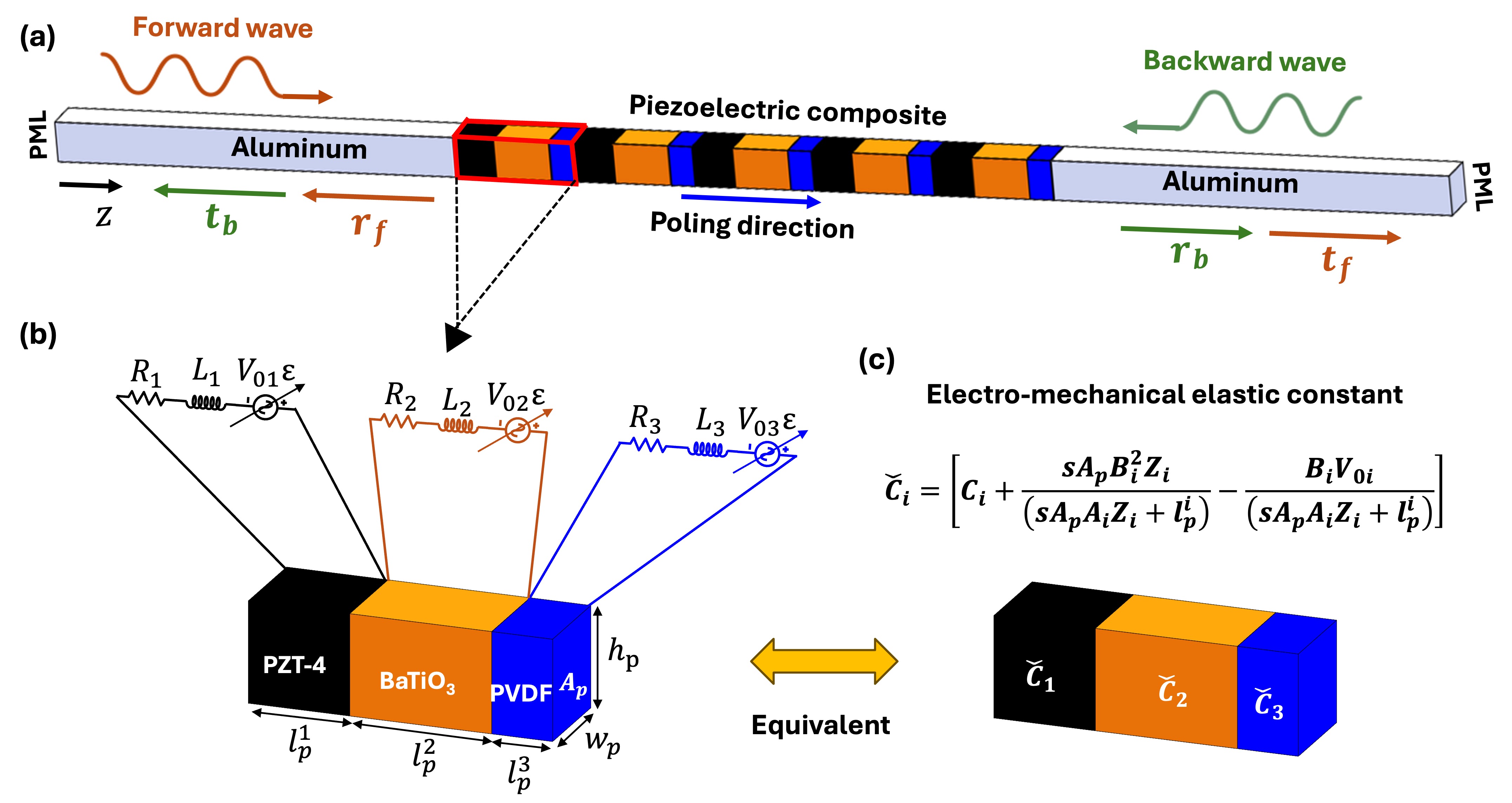}
    \caption{\textbf{Schematic and modeling framework for electro-momentum coupling in shunted piezoelectric metamaterials.}
    \textbf{a}, One-dimensional waveguide comprising a layered piezoelectric composite embedded between two aluminum sections and terminated with perfectly matched layers (PML). Elastic waves incident from either direction exhibit asymmetric transmission ($t_f$, $t_b$) and reflection ($r_f$, $r_b$) due to the spatial asymmetry introduced by the shunted composite. The structure features a uniform poling direction, while the arrangement of heterogeneous piezoelectric layers breaks inversion symmetry to support directional wave propagation. 
    \textbf{b}, Zoomed-in view of a representative unit cell consisting of three serially connected piezoelectric layers PZT-4, BaTiO$_3$, and PVDF each interfaced with a shunt circuit composed of a resistor ($R_i$), inductor ($L_i$), and strain-proportional voltage feedback source ($V_{0i} \varepsilon$). These circuits actively tailor the electromechanical response of each segment and modulate the macroscopic dynamic behavior through piezoelectric coupling. 
    \textbf{c}, The shunted layer is modeled as an effective elastic layer with a modified electro-mechanical elastic constant $\check{C}_i$, incorporating both material properties and circuit-induced effects. This framework enables tunable control of wave propagation through circuit parameters, where the resistor introduces loss, the inductor defines resonance behavior, and the voltage feedback allows dynamic modulation of stiffness.}
    \label{fig:shunt_schematic}
\end{figure}

We consider wave propagation through a finite, one-dimensional periodic piezoelectric composite consisting of layered piezoelectric materials stacked along their poling direction. The schematic in Fig.~\ref{fig:shunt_schematic}(a) shows a finite rod composed of five unit cells of the piezoelectric composite embedded within an aluminum rod. Wave propagation is studied in both forward and backward directions by analyzing the reflected and transmitted wave components for an incident wave in each direction. Figure~\ref{fig:shunt_schematic}(b) illustrates the unit cell of the composite, which comprises piezoelectric layers of PZT-4, BaTiO$_3$, and PVDF. These materials are stacked in a specific pattern to break inversion symmetry, enabling the realization of Willis and electro-momentum coupling and thereby facilitating asymmetric wave phenomena. All layers have the same cross-sectional area, $A_p = 1\,\text{mm}^2$, while their lengths are selected as $l_p^1 = 1\,\text{mm}$, $l_p^2 = 1.4\,\text{mm}$, and $l_p^3 = 0.6\,\text{mm}$.

Each layer is shunted with an external circuit consisting of an inductor, a resistor, and a voltage feedback source proportional to the strain across the layer. An effective electro-mechanical elastic constant is derived, allowing the shunted piezoelectric layers to be modeled as an equivalent purely elastic medium. The influence of piezoelectric coupling and circuit parameters is incorporated into the modified constitutive behavior, as illustrated in Fig.~\ref{fig:shunt_schematic}(c). In this formulation, \( \check{C} \) denotes the electro-mechanical elastic constant, which captures the combined effects of the intrinsic material properties and the external circuitry. Here, \( C \) and \( A \) represent the elastic and dielectric constants, respectively, and \( B \) is the piezoelectric coupling coefficient. The term \( V_0 \) is the voltage feedback coefficient, and \( Z = sL + R \) is the shunt impedance, where \( L \) and \( R \) are the inductance and resistance of the circuit, respectively. The Laplace variable is defined as \( s = -i \omega \), with \( \omega \) being the angular frequency. 

The resistor introduces damping into the system, while the inductor forms an LC circuit resonance with the intrinsic capacitance of the piezoelectric layer. The voltage feedback source provides additional control over the dynamic behavior by exerting an active response proportional to the strain across the layer. Previously, we demonstrated that the resistor and inductor in the external circuit can be used to control Willis and electro-momentum coupling, thereby tuning asymmetric wave phenomena by adjusting these circuit parameters~\cite{danawe2023electro}. Specifically, the LC resonance determines the frequency at which coupling effects are maximized, while the resistor governs the perturbation level of the asymmetry factor, which directly influences the degree of wave asymmetry. By introducing a voltage feedback source proportional to strain, we now demonstrate an enhanced level of control that enables the realization of exotic wave phenomena such as unidirectional zero reflection and perfect absorption.

\subsection*{Scattering analysis}

\begin{figure}[ht]
    \centering
    \includegraphics[width=\linewidth]{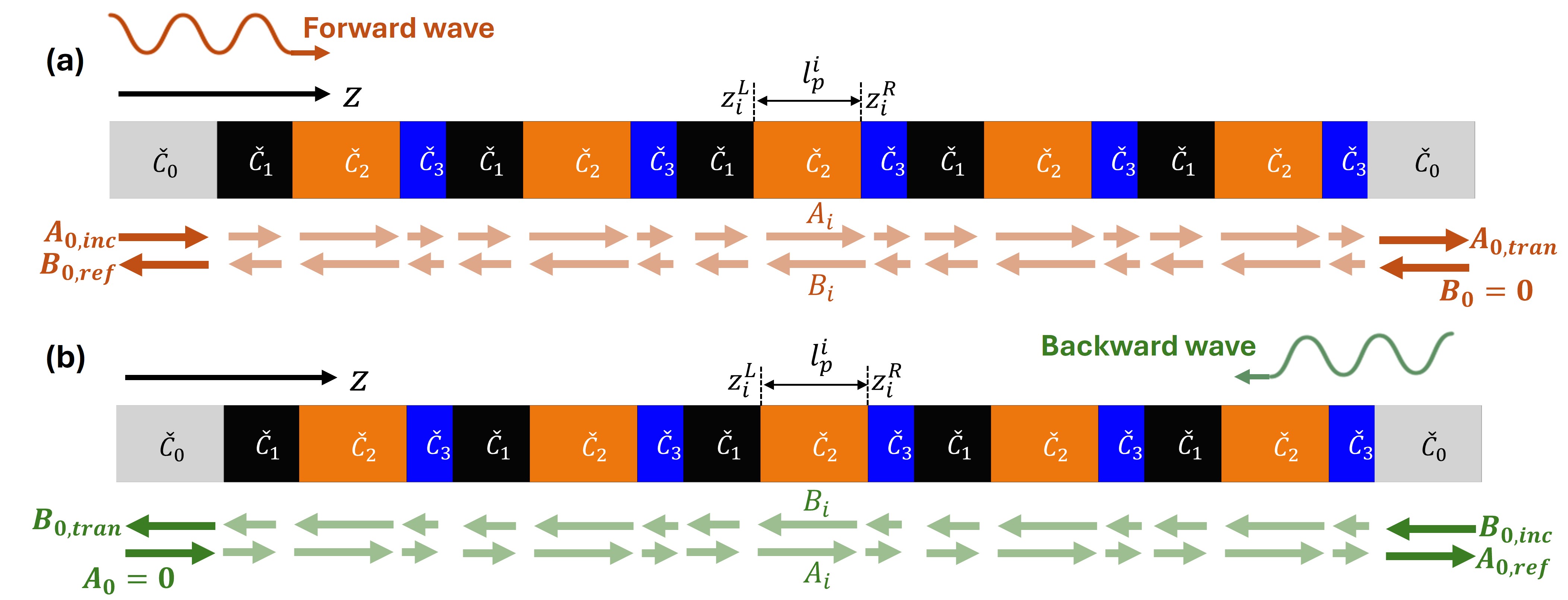}
    \caption{\textbf{Wave decomposition in a layered piezoelectric composite for forward and backward incidence.}
    \textbf{a}, Schematic of elastic wave propagation from the left (forward incidence) through a layered structure composed of alternating piezoelectric segments with effective electro-mechanical elastic constants $\check{C}_1$, $\check{C}_2$, and $\check{C}_3$, bounded by homogeneous elastic layers $\check{C}_0$. Each segment is associated with local forward ($A_i$) and backward ($B_i$) wave components. The incident wave ($A_{0,\text{inc}}$) is partially reflected ($B_{0,\text{ref}}$) and transmitted ($A_{0,\text{tran}}$), with no incoming wave from the right ($B_0 = 0$). 
    \textbf{b}, The same configuration under backward incidence, where the wave enters from the right. The incident wave ($B_{0,\text{inc}}$) results in partial transmission ($B_{0,\text{tran}}$) and reflection ($A_{0,\text{ref}}$), with no incoming wave from the left ($A_0 = 0$). This layer-wise decomposition into forward and backward components enables analytical evaluation of reflection and transmission coefficients under both excitation directions using the transfer matrix method.}
    \label{fig:transfer_matrix}
\end{figure}

The asymmetric wave behavior is analyzed using a scattering matrix composed of transmission and reflection ratios for forward and backward wave incidence. This formulation is based on the standard transfer matrix method, which relates the wave amplitudes across each layer of the piezoelectric composite. Figure~\ref{fig:transfer_matrix}(a) and (b) illustrate the traveling wave components within each layer for forward and backward incidence, respectively, along with the incident (\( A_{0,\mathrm{inc}} \), \( B_{0,\mathrm{inc}} \)), reflected (\( B_{0,\mathrm{ref}} \), \( A_{0,\mathrm{ref}} \)), and transmitted (\( A_{0,\mathrm{trans}} \), \( B_{0,\mathrm{trans}} \)) wave amplitudes in the surrounding aluminum rod. These field distributions form the foundation for computing direction-dependent scattering parameters that reveal the emergence of wave asymmetry induced by circuit-modulated electro-mechanical coupling in the composite.

Each layer is modeled using an effective elastic constant. For the aluminum segments, this corresponds to the intrinsic elastic constant (i.e., \( \check{C}_0 = C_0 \)), while for the piezoelectric layers, it represents an electro-mechanical elastic constant \( \check{C}_i \), which accounts for both the intrinsic material response and the effects of the external shunt circuitry, as described in Fig.~\ref{fig:shunt_schematic}(c). 
The reflection and transmission coefficients are defined as the ratios of outgoing to incoming wave amplitudes:
\begin{align}
r_f &= \frac{B_{0,\mathrm{ref}}}{A_{0,\mathrm{inc}}}, \quad 
r_b = \frac{A_{0,\mathrm{ref}}}{B_{0,\mathrm{inc}}} \nonumber \\
t_f &= \frac{A_{0,\mathrm{trans}} \, e^{i k_0 \ell}}{A_{0,\mathrm{inc}}}, \quad
t_b = \frac{B_{0,\mathrm{trans}} \, e^{i k_0 \ell}}{B_{0,\mathrm{inc}}}
\end{align}
where \( A_{0,\mathrm{inc}} \) and \( B_{0,\mathrm{inc}} \) are the amplitudes of incident waves in the forward and backward directions, respectively. \( B_{0,\mathrm{ref}} \) and \( A_{0,\mathrm{ref}} \) denote reflected waves, and \( A_{0,\mathrm{trans}} \), \( B_{0,\mathrm{trans}} \) are the transmitted waves. The exponential term \( e^{i k_0 \ell} \) accounts for the phase accumulated over the length \( \ell \) of the composite, with \( k_0 \) as the wavenumber in the aluminum background.

\begin{figure}[!t]
    \centering
    \includegraphics[width=\linewidth]{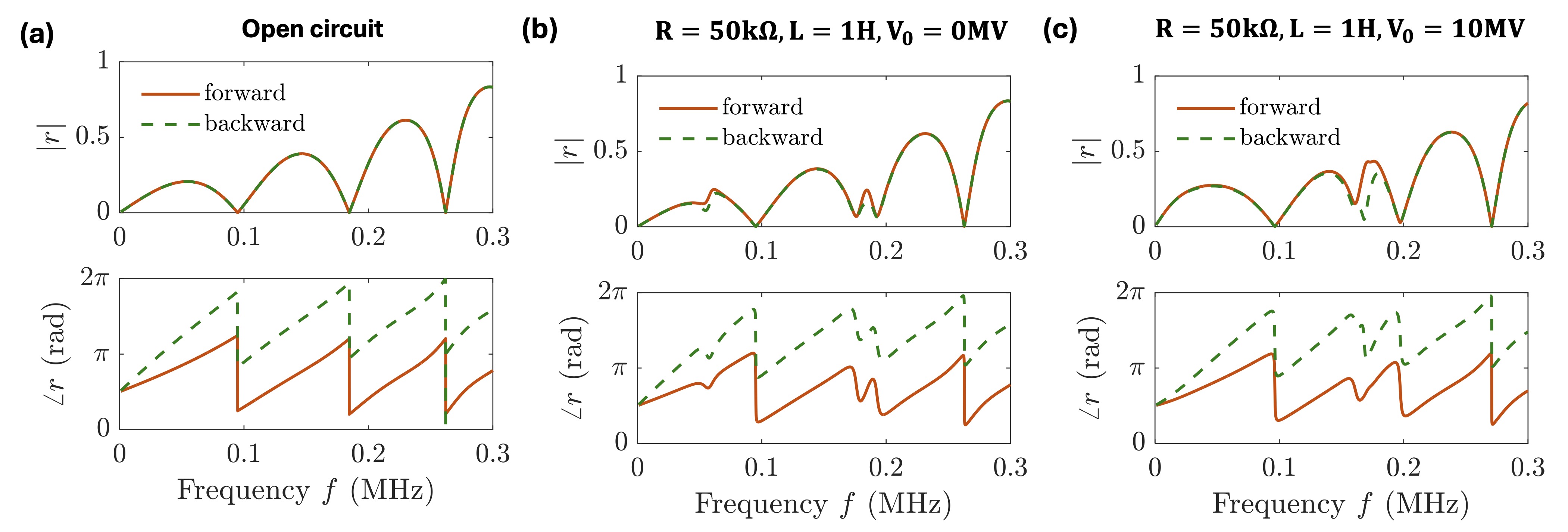}
    \caption{\textbf{Effect of shunt circuit parameters on directional reflection behavior.} 
    Frequency-dependent reflection magnitude ($|r|$, top row) and phase ($\angle r$, bottom row) for forward (solid lines) and backward (dashed lines) wave incidence are shown under different shunting conditions. Identical circuit parameters are applied to all piezoelectric layers.  
    \textbf{a}, In the open-circuit case (no shunt), the reflection magnitudes are symmetric, while phase asymmetry arises solely due to structural asymmetry. 
    \textbf{b}, Introducing a resistor ($R = 50~\mathrm{k}\Omega$) and inductor ($L = 1~\mathrm{H}$) leads to asymmetry in both magnitude and phase of the reflection ratio, with losses induced by the resistor breaking amplitude symmetry. The LC resonance condition contributes to frequency-selective enhancement of this asymmetry. 
    \textbf{c}, Adding voltage feedback ($V_0 = 10~\mathrm{MV}$) further modifies the reflection characteristics across the entire frequency spectrum, introducing tunability and stronger contrast between forward and backward responses. These results illustrate how circuit parameters enable directional control over elastic wave reflection in shunted piezoelectric composites.}
    \label{fig:reflection_ratio}
\end{figure}

 The expressions for reflection and transmission ratios are:
\begin{align}
r_f &= -\frac{\mathbf{M}_f(2,1)}{\mathbf{M}_f(2,2)}, \quad
t_f = \mathbf{M}_f(1,1) - \mathbf{M}_f(1,2)\frac{\mathbf{M}_f(2,1)}{\mathbf{M}_f(2,2)} \nonumber \\
r_b &= -\frac{\mathbf{M}_b(2,1)}{\mathbf{M}_b(2,2)}, \quad
t_b = \mathbf{M}_b(1,1) - \mathbf{M}_b(1,2)\frac{\mathbf{M}_b(2,1)}{\mathbf{M}_b(2,2)}
\end{align} where $\mathbf{M}_f$ and $\mathbf{M}_b$ are given by equations \ref{eq:Mf} and \ref{eq:Mb}, respectively. The magnitude and phase of the reflection coefficients for both forward and backward incidence are plotted in Fig.~\ref{fig:reflection_ratio}(a)–(c) for three different circuit configurations, where the circuit parameters inductance, resistance, and voltage feedback are identical across all three layers within each unit cell. In the open-circuit case [Fig.~\ref{fig:reflection_ratio}(a)], the reflection amplitude remains symmetric, with only phase asymmetry observed between the two directions. Introducing a passive shunt with \( R = 50\,\text{k}\Omega \) and \( L = 1\,\text{H} \) [Fig.~\ref{fig:reflection_ratio}(b)] produces measurable asymmetry in both amplitude and phase, particularly near the LC resonance frequencies, \( f_1 = 0.067\,\mathrm{MHz} \) for the PZT-4 layer and \( f_2 = 0.19\,\mathrm{MHz} \) for the BaTiO$_3$ layer. The LC resonance frequency of the PVDF layer lies beyond the frequency range considered in this study and is therefore not captured in the plotted results. The amplitude asymmetry primarily arises from energy dissipation introduced by the resistor, which leads to direction-dependent wave attenuation. While dissipation alone does not fundamentally break time-reversal symmetry, it facilitates asymmetric scattering when combined with structural or parametric asymmetries, as is the case in this configuration.

Finally, incorporating an active voltage feedback source with \( V_0 = 10\,\text{MV} \) [Fig.~\ref{fig:reflection_ratio}(c)] results in pronounced asymmetry, where both the magnitude and phase of the reflection coefficients differ significantly between forward and backward wave incidence. These effects are especially prominent near the frequencies where the circuit dynamics enhance Willis and electromomentum coupling. These results demonstrate that external circuit parameters, including both passive and active elements, provide a powerful means to control and tune asymmetric wave behavior. In particular, the emergence of strong reflection asymmetry underscores the potential of dynamic electro-mechanical coupling for breaking reciprocity in piezoelectric composite systems.

\subsection*{Exceptional points and unidirectional zero reflection}

Unidirectional zero reflection (UZR) can be realized at exceptional points (EPs), where the eigenvalues and eigenvectors of the scattering matrix coalesce. The scattering matrix \( \mathbf{S} \) is defined as:
\begin{align}
\label{eq:scattering_matrix}
\mathbf{S} &=
\begin{bmatrix}
t_f & r_b \\
r_f & t_b
\end{bmatrix}, \quad \mathrm{eig}(\mathbf{S}) = \lambda_1, \lambda_2 \nonumber \\
\lambda_1 &= t + \sqrt{r_f r_b}, \quad \lambda_2 = t - \sqrt{r_f r_b}
\end{align}

\begin{figure}[!t]
    \centering
    \includegraphics[width=\linewidth]{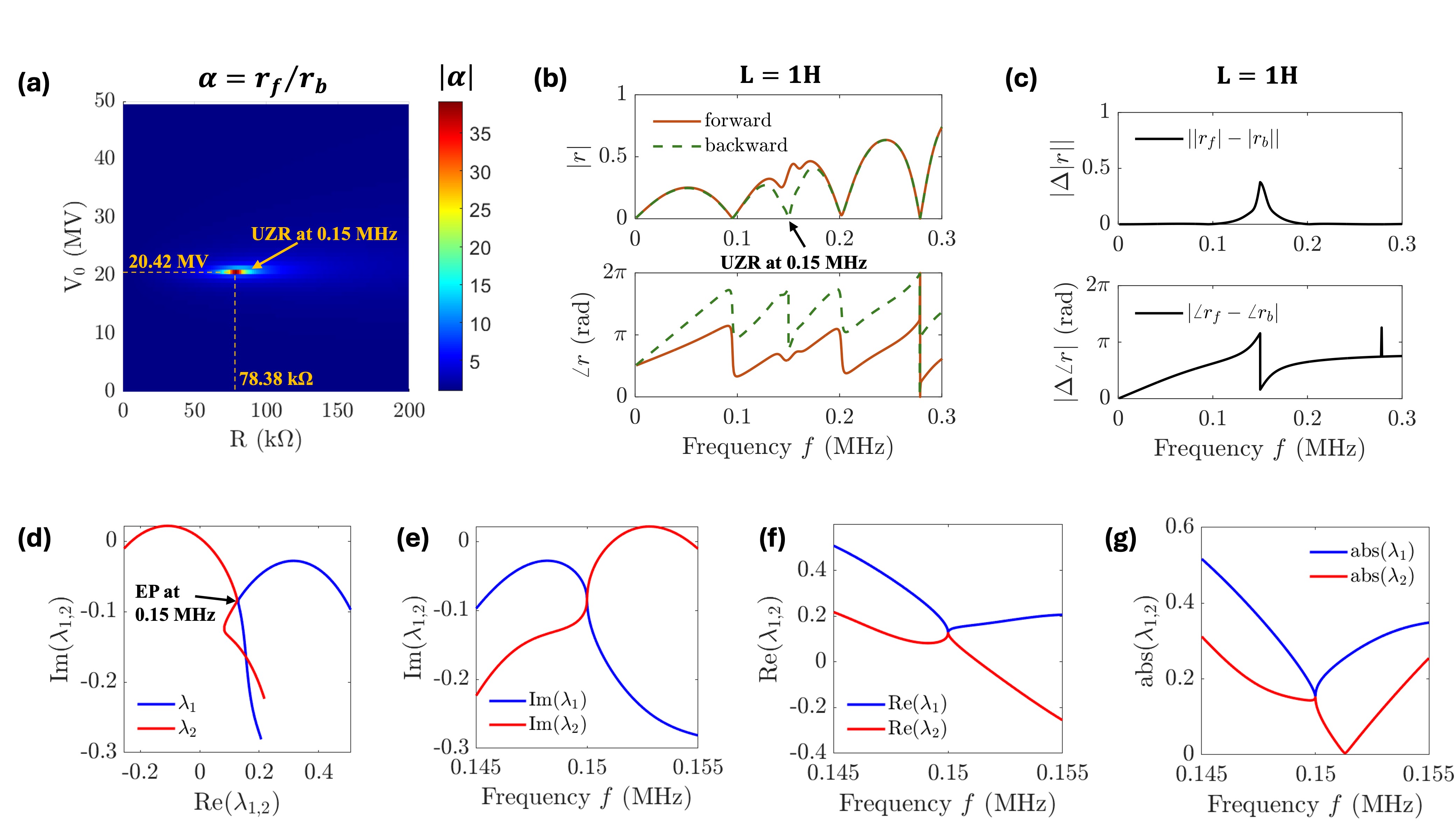}
    \caption{\textbf{Parametric analysis of unidirectional zero reflection (UZR) and its relation to exceptional points (EPs).} 
    \textbf{a}, Colormap of the contrast ratio $\alpha = r_f / r_b$ between forward and backward reflection as a function of resistance ($R$) and voltage feedback amplitude ($V_0$), with inductance fixed at $L = 1~\mathrm{H}$. Identical circuit parameters are applied to all piezoelectric layers. A sharp peak in $\alpha$ identifies the condition for UZR at $0.15~\mathrm{MHz}$, corresponding to $R = 78.38~\mathrm{k}\Omega$ and $V_0 = 20.42~\mathrm{MV}$.
    \textbf{b}, Frequency-dependent reflection magnitude (top) and phase (bottom) for forward (solid) and backward (dashed) incidence, confirming complete suppression of forward reflection at the UZR frequency along with a $\pi$ phase difference. 
    \textbf{c}, Absolute differences in reflection magnitude (top) and phase (bottom) between forward and backward incidence, showing a pronounced asymmetry at $0.15~\mathrm{MHz}$. 
    \textbf{d}, Eigenvalue trajectories of the scattering matrix in the complex plane showing the coalescence of $\lambda_1$ and $\lambda_2$ at the EP.
    \textbf{e}–\textbf{g}, Frequency-dependent plots of the imaginary parts (\textbf{e}), real parts (\textbf{f}), and magnitudes (\textbf{g}) of the eigenvalues, confirming the occurrence of a second-order exceptional point at the UZR frequency. The $\pi$ phase jump observed in \textbf{b} and \textbf{c} is a spectral signature of the EP. These results confirm that UZR coincides with an exceptional point of the scattering matrix composed of reflection and transmission coefficients ($r_f$, $r_b$, $t_f$, $t_b$).}
    \label{fig:UZR_EP}
\end{figure} Here, \( \lambda_1 \) and \( \lambda_2 \) are the eigenvalues of the scattering matrix. Since the system is reciprocal, the transmission is equal in both directions, i.e., \( t = t_f = t_b \). At the UZR condition, either \( r_f = 0 \) or \( r_b = 0 \), resulting in \( \lambda_1 = \lambda_2 = t \), which indicates a coalescence of eigenvalues and eigenvectors, a defining signature of an exceptional point. The scattering matrix elements \( t_{f,b} \) and \( r_{f,b} \), as derived in Eq.~(9), can be used to locate EPs in the system. By tuning the external circuit parameters, these non-Hermitian degeneracies can be accessed, enabling UZR behavior in the piezoelectric metamaterial.\begin{figure}[!b]
    \centering
    \includegraphics[width=\linewidth]{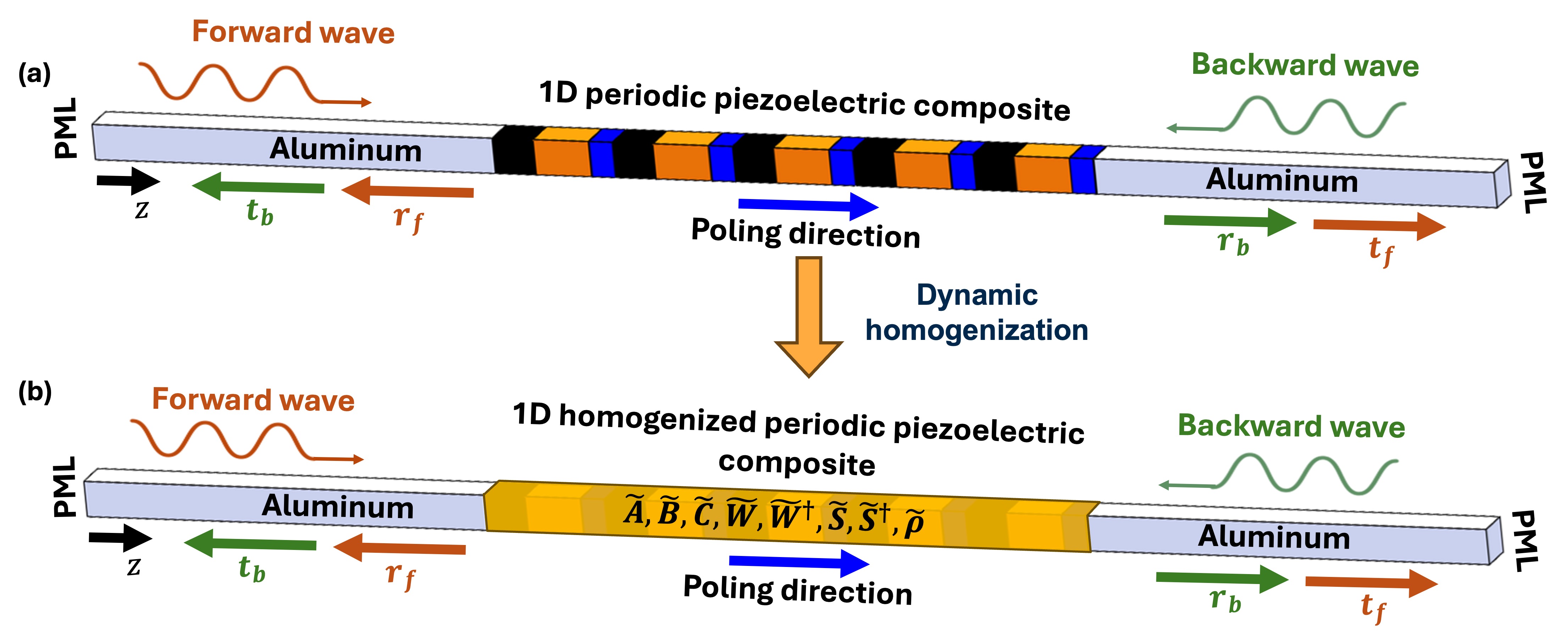}
    \caption{\textbf{Dynamic homogenization of a layered piezoelectric composite.}
    \textbf{a}, Schematic of a one-dimensional periodic piezoelectric composite consisting of alternating heterogeneous piezoelectric layers embedded between aluminum waveguide sections and terminated with perfectly matched layers (PML). Due to the lack of inversion symmetry and spatial variation, elastic waves exhibit direction-dependent reflection ($r_f$, $r_b$). 
    \textbf{b}, The composite is dynamically homogenized into an equivalent continuum domain characterized by effective macroscopic properties. These include the stiffness tensor $\tilde{\mathbf{C}}$, effective density $\tilde{\mathbf{\rho}}$, and piezoelectric coupling tensors $\tilde{\mathbf{A}}$ and $\tilde{\mathbf{B}}$. Importantly, the homogenized model also incorporates the Willis coupling tensors $\tilde{\mathbf{S}}$, $\tilde{\mathbf{S}}^\dagger$ (describing coupling between momentum and strain), and the electro-momentum coupling tensors $\tilde{\mathbf{W}}$, $\tilde{\mathbf{W}}^\dagger$ (capturing coupling between electric field and momentum). The homogenized medium retains the same poling direction as the original layered system and reproduces its macroscopic wave behavior. This framework enables effective modeling of asymmetric wave propagation in complex piezoelectric composites.}
    \label{fig:homogenization}
\end{figure}To systematically identify UZR conditions, we perform a parametric sweep over resistance \( R \) and feedback voltage \( V_0 \), while keeping the inductance fixed at \( L = 1\,\mathrm{H} \). The circuit parameters are kept identical across all three layers of the unit cell. A contrast factor is defined to quantify asymmetry:
\begin{equation}
\alpha = \frac{r_f}{r_b}
\end{equation}

At UZR, \( |\alpha| \rightarrow 0 \) or \( \infty \). Figure~\ref{fig:UZR_EP}(a) presents the magnitude of \( \alpha \) as a function of \( R \) and \( V_0 \) at a design frequency of \( 0.15\,\mathrm{MHz} \), revealing a region of strong reflection asymmetry. A distinct UZR point is observed at \( R = 78.38\,\mathrm{k}\Omega \), \( V_0 = 20.42\,\mathrm{MV} \), corresponding to an EP in the system. The reflection coefficients at this parameter combination are plotted in Fig.~\ref{fig:UZR_EP}(b), confirming the suppression of backward reflection \( (r_b \approx 0) \) and a large contrast between forward and backward response.

To further quantify this asymmetry, the differential amplitude \( ||r_f| - |r_b|| \) and differential phase \( |\angle r_f - \angle r_b| \) are plotted in Fig.~\ref{fig:UZR_EP}(c). A peak in the amplitude difference and a sharp phase jump of nearly \( \pi \) confirm the asymmetric scattering characteristic of an EP. The presence of the exceptional point is further validated by examining the eigenvalue evolution of the scattering matrix. In Fig.~\ref{fig:UZR_EP}(d), the eigenvalues are plotted in the complex plane, showing a coalescence trajectory near the EP. Figures~\ref{fig:UZR_EP}(e)–(g) present the real, imaginary, and absolute parts of the eigenvalues as functions of frequency. At the UZR frequency, both real and imaginary components coalesce, confirming the degeneracy of eigenvalues and reinforcing the presence of a non-Hermitian EP.
\begin{figure}[!b]
    \centering
    \includegraphics[width=\linewidth]{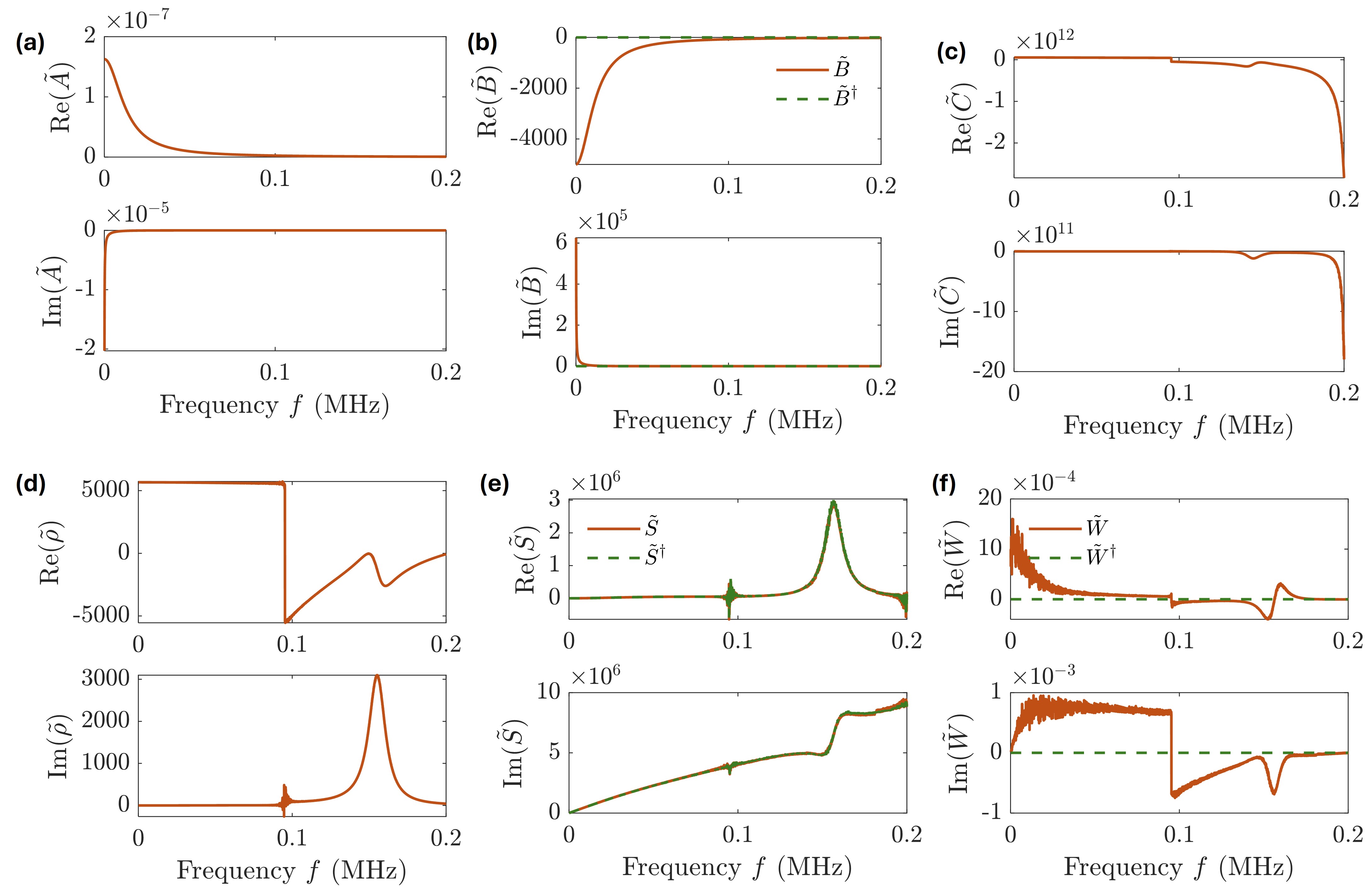}
    \caption{\textbf{Effective constitutive properties of the layered piezoelectric composite exhibiting UZR at 0.15\,MHz.}
    Frequency-dependent homogenized parameters are extracted using dynamic transfer matrix–based homogenization for the circuit configuration yielding unidirectional zero reflection (UZR). 
    \textbf{a}, Dielectric stiffness \( \tilde{A} \). 
    \textbf{b}, Piezoelectric coupling \( \tilde{B} \) and its Hermitian conjugate \( \tilde{B}^\dagger \) deviate significantly, confirming broken symmetry. 
    \textbf{c}, Electro-mechanical stiffness \( \tilde{C} \) reflects circuit-tuned stiffness modulation. 
    \textbf{d}, Effective density \( \tilde{\rho} \) shows a discontinuity and complex-valued behavior near 0.15\,MHz, indicating resonance. 
    \textbf{e}, Willis coupling \( \tilde{S} \) and its symmetric off-diagonal term \( \tilde{S}^{\dagger} \) demonstrate non-Hermitian asymmetry, peaking near the UZR condition. 
    \textbf{f}, Electro-momentum coupling \( \tilde{W} \) and \( \tilde{W}^{\dagger} \)confirm the emergence of non-Hermitian behavior.}
    \label{fig:Effective_Properties}
\end{figure}

\begin{figure}[!b]
    \centering
    \includegraphics[width=\linewidth]{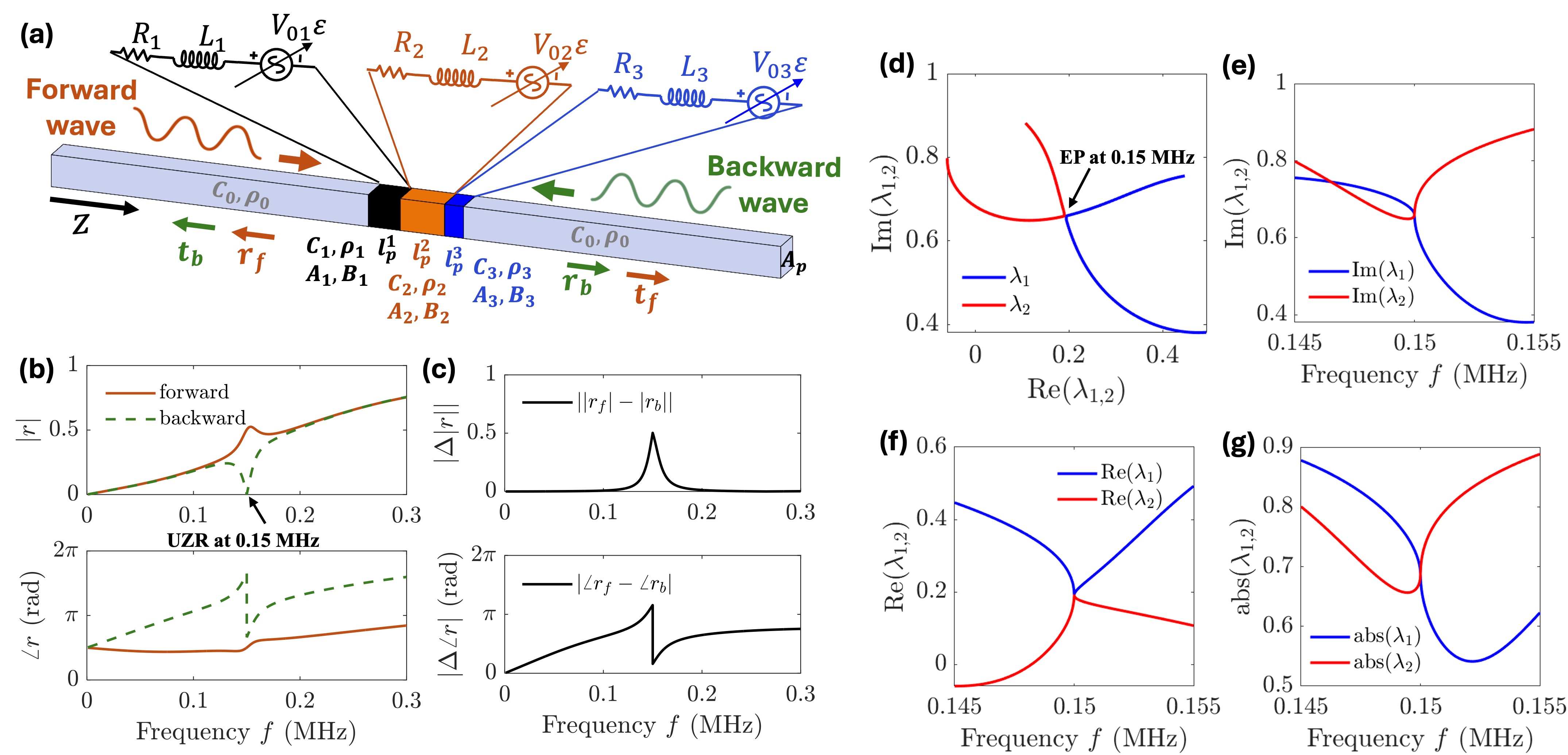}
    \caption{\textbf{Unidirectional zero reflection and exceptional point behavior in a single-unit-cell piezoelectric composite.} 
    \textbf{a}, Schematic of a one-dimensional waveguide consisting of a single unit cell made of three piezoelectric layers (PZT-4, BaTiO$_3$, and PVDF), each connected to an identical shunt circuit comprising a resistor ($R_i$), inductor ($L_i$), and strain-proportional voltage feedback source ($V_{0i}\varepsilon$). The waveguide is bounded by homogeneous aluminum sections and terminated with perfectly matched layers (PML).  
    \textbf{b}, Reflection magnitude (top) and phase (bottom) for forward (solid) and backward (dashed) incidence show a strong asymmetry near $0.15~\mathrm{MHz}$, where forward reflection vanishes, indicating unidirectional zero reflection (UZR).  
    \textbf{c}, Difference in reflection magnitude ($|\Delta r|$) and phase ($|\Delta \angle r|$) between forward and backward directions, both peaking at the UZR frequency.  
    \textbf{d}, Trajectory of the scattering matrix eigenvalues $\lambda_1$ and $\lambda_2$ in the complex plane, indicating coalescence at an exceptional point (EP).  
    \textbf{e}–\textbf{g}, Frequency-dependent evolution of the imaginary parts (\textbf{e}), real parts (\textbf{f}), and absolute values (\textbf{g}) of the eigenvalues further confirm the occurrence of a second-order EP at $0.15~\mathrm{MHz}$.  
    These results demonstrate that both UZR and EP phenomena can be realized with just a single unit cell, underscoring the strong asymmetry induced by shunt-circuit-mediated electro-momentum coupling.}
    \label{fig:UZR_EP_singleUC}
\end{figure}

These results demonstrate that external circuit tuning via resistive, inductive, and voltage feedback elements provides a powerful and reconfigurable platform for accessing exceptional points and achieving UZR in elastic metamaterials. Unlike conventional systems where EPs are fixed by material properties or geometry, the piezoelectric-circuit hybrid structure enables on-demand tuning of asymmetric wave transport. This approach opens new directions for reprogrammable waveguides, vibration control, and logic devices based on elastic waves.

\subsection*{Dynamic Homogenization}
The effective properties of a heterogeneous layered composite, derived via dynamic homogenization~\cite{pernas2020symmetry}, can capture unusual coupling mechanisms that arise from broken inversion symmetry in periodic systems. Among these, Willis coupling and electro-momentum coupling play a central role in enabling asymmetric wave propagation. In our previous work~\cite{danawe2023electro}, we employed an ensemble-averaging approach to extract the effective constitutive parameters of an infinitely periodic piezoelectric composite with external shunt circuits in the long-wavelength limit. This framework allowed us to rigorously quantify the influence of a passive shunt circuit, comprising an inductor and a resistor, on the effective medium properties. In particular, we demonstrated that the inclusion of such circuits modifies the asymmetry factor, a key parameter that governs the degree of wave asymmetry.

In this study, we adopt a transfer matrix-based homogenization approach\cite{huynh2025effect} to replace the finite piezoelectric layered system with a single homogenized domain, as illustrated in Fig.~\ref{fig:homogenization}. Figure~\ref{fig:homogenization}(a) depicts the original heterogeneous structure, where wave scattering is computed using the microscopic properties of individual layers. In contrast, dynamic homogenization yields an effective medium, as shown in Fig.~\ref{fig:homogenization}(b), characterized by emergent coupling terms that do not appear in conventional media. Specifically, the effective constitutive relation reveals additional Willis coupling terms, $\tilde{S}, \tilde{S}^{\dagger}$, and electro-momentum coupling terms, $\tilde{W}, \tilde{W}^{\dagger}$, such that the generalized 1D constitutive law is given by:
\begin{equation}
\label{eq:Effective_Const_Law}
\begin{pmatrix}
\langle \sigma \rangle \\
\langle D \rangle \\
\langle p \rangle
\end{pmatrix}
=
\begin{pmatrix}
\tilde{C} & -\tilde{B} & \tilde{S} \\
\tilde{B} & \tilde{A} & \tilde{W} \\
\tilde{S}^{\dagger} & -\tilde{W}^{\dagger} & \tilde{\rho}
\end{pmatrix}
\begin{pmatrix}
\langle \varepsilon \rangle \\
\langle E \rangle \\
\langle \dot{u} \rangle
\end{pmatrix}
\end{equation} where all quantities are scalars representing spatially averaged effective fields: $\langle \sigma \rangle$ is the stress, $\langle D \rangle$ the electric displacement, and $\langle p \rangle$ the momentum density, while $\langle \varepsilon \rangle$, $\langle E \rangle$, and $\langle \dot{u} \rangle$ denote the strain, electric field, and particle velocity, respectively. The coefficients $\tilde{C}$, $\tilde{A}$, and $\tilde{\rho}$ represent the effective stiffness, permittivity, and mass density; $\tilde{B}$ is the piezoelectric coupling; and $\tilde{S}$,$\tilde{S}^{\dagger}$ and $\tilde{W}$,$\tilde{W}^{\dagger}$ encode the Willis and electro-momentum couplings. To extract the effective properties, a transfer matrix is constructed for both the heterogeneous and homogenized representations of the piezoelectric composite. These matrices are compared by matching their scattering coefficients under identical boundary and loading conditions. Since the homogenized constitutive relation (Eq.~\eqref{eq:Effective_Const_Law}) contains nine unknown parameters, the transfer matrix must be defined to fully capture both mechanical and electrical degrees of freedom.

For the heterogeneous composite, we define a four-dimensional state vector that includes the mechanical displacement \( u \), electric potential \( \phi \), stress \( \sigma \), and electric displacement \( D \). Under time-harmonic excitation, the governing equations are written in first-order form as:
\begin{equation}
\label{eq:State_Vector_EOM}
\frac{d}{dz}
\begin{pmatrix}
u \\
\phi \\
\sigma \\
D
\end{pmatrix}
=
\begin{bmatrix}
0 & 0 & \dfrac{1}{\check{C}} & 0 \\
0 & 0 & -\dfrac{V_0}{\check{C} \, l_p} & \dfrac{s A_p Z}{l_p} \\
s^2 \rho & 0 & 0 & 0 \\
0 & 0 & 0 & 0
\end{bmatrix}
\begin{pmatrix}
u \\
\phi \\
\sigma \\
D
\end{pmatrix}
\end{equation} where, \( \rho \) is the mass density, and all other parameters are as previously defined. This formulation enables the construction of a \( 4 \times 4 \) transfer matrix that accurately captures the coupled electromechanical behavior of the heterogeneous piezoelectric unit cell. It accounts for the interaction between the mechanical and electrical fields mediated by the piezoelectric effect, as well as the influence of the external shunt circuit through the impedance \( Z \) and feedback coefficient \( V_0 \). Refer to the methods section for the derivation of the transfer matrix and effective properties. To elucidate the physical origins of asymmetric scattering behavior, we compute the frequency-dependent effective constitutive properties of the layered piezoelectric composite using a dynamic homogenization approach. The configuration corresponds to the parameter set obtained from the earlier parametric study, which exhibited unidirectional zero reflection (UZR) at 0.15\,MHz.

Figure~\ref{fig:Effective_Properties} shows the real and imaginary components of the homogenized constitutive properties. As evident in these plots, the composite exhibits strong frequency-dependent non-Hermiticity due to the lossy and active components in the shunt circuits, as reflected in the large imaginary parts of the constitutive tensors. In particular, the emergence of non-zero \( \tilde{S} \) and \( \tilde{W} \), which break reciprocity and time-reversal symmetry, confirms the presence of Willis and electro-momentum coupling enabled by the broken inversion symmetry and external circuit feedback. These couplings play a central role in generating asymmetric wave reflection.

Importantly, the Hermitian conjugates of the off-diagonal coupling terms \( \tilde{B}^\dagger \) and \( \tilde{W}^\dagger \) (shown as dashed lines) deviate from the original tensors, confirming the non-Hermitian nature of the effective medium. This validates that the asymmetric wave propagation arises from both geometric asymmetry and circuit-induced gain/loss modulation. Notably, a sharp discontinuity in \( \tilde{\rho} \) and resonance peaks in \( \tilde{S} \) and \( \tilde{W} \) coincide with the UZR frequency at 0.15\,MHz, highlighting the critical role of dynamic resonances in tuning wave asymmetry.

\subsection*{Unidirectional zero reflection and perfect absorption}

Consider a configuration with a single unit cell embedded in an aluminum background rod, as illustrated in Fig.~\ref{fig:UZR_EP_singleUC}(a). Based on the parametric study conducted earlier with a five–unit cell composite piezoelectric rod (see Fig.~\ref{fig:UZR_EP}), UZR is achieved at \( R = 78.38\,\text{k}\Omega \) and \( V_0 = 20.42\,\text{MV} \), corresponding to an EP when the inductance is set to \( L = 1\,\text{H} \). Remarkably, this condition also holds for a single scatterer, as verified through a scattering analysis and reflection measurements.

The persistence of UZR behavior in a single unit cell stems from the local nature of the non-Hermitian scattering mechanism. Specifically, the piezoelectric layer shunted with an external circuit introduces an effective non-Hermitian coupling that governs the reflection asymmetry. Since the scattering matrix for this configuration remains a \( 2 \times 2 \) system, the coalescence of its eigenvalues (i.e., the EP condition) and the associated unidirectional response can emerge even in the absence of periodicity. Thus, the essential features of UZR and EP behavior are preserved in the single-cell configuration when the circuit parameters are properly tuned. Figure~\ref{fig:UZR_EP_singleUC}(b) presents the frequency-dependent reflection amplitude and phase for both forward and backward incidence. UZR is observed at 0.15\,MHz, where the reflection magnitude for backward incidence vanishes. The corresponding reflection asymmetry, quantified in Fig.~\ref{fig:UZR_EP_singleUC}(c), shows a sharp peak in amplitude contrast and a phase difference approaching \( \pi \), which is a characteristic signature of an EP. Further confirmation is provided in Figs.~\ref{fig:UZR_EP_singleUC}(d)–(g), where the eigenvalues of the scattering matrix extracted from the single-cell response are shown to coalesce at the EP frequency. The trajectory in the complex plane [Fig.~\ref{fig:UZR_EP_singleUC}(d)], along with the frequency evolution of their imaginary, real, and absolute components [Figs.~\ref{fig:UZR_EP_singleUC}(e–g)], collectively validates the existence of an exceptional point in the reduced system. \begin{figure}[!b]
    \centering
    \includegraphics[width=\linewidth]{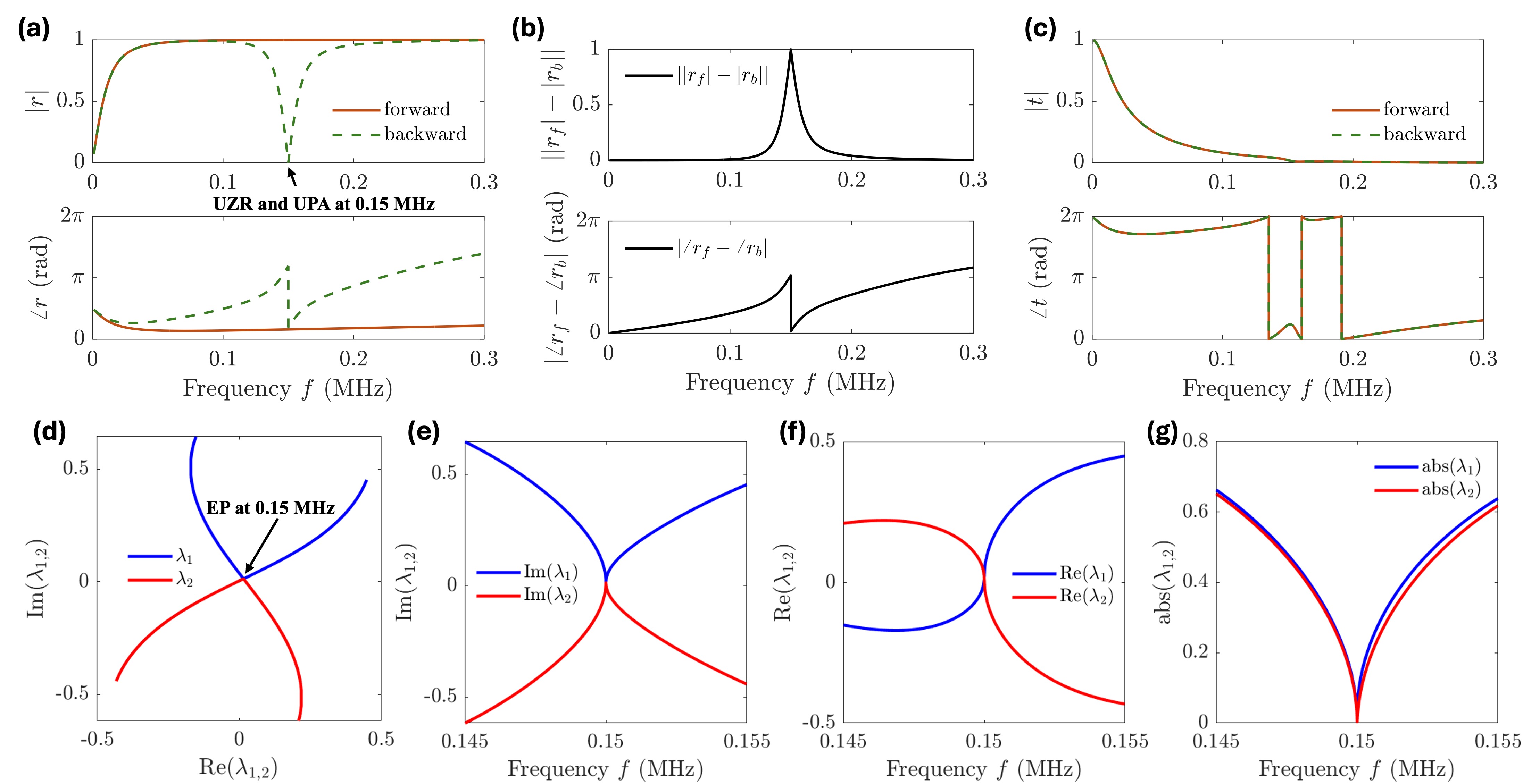}
    \caption{\textbf{Simultaneous unidirectional perfect absorption and zero reflection via multi-parameter circuit optimization.}
    (a) Reflection magnitude (top) and phase (bottom) for forward (solid) and backward (dashed) incidence. At 0.15 MHz, the system exhibits near-unity reflection for forward incidence and near-zero reflection for backward incidence. As transmission is also negligible in both directions, this results in perfect absorption in the backward direction (zero reflection and zero transmission) and perfect reflection in the forward direction (unity reflection with no transmission or absorption).
    (b) Reflection asymmetry in both magnitude and phase between forward and backward incidence, peaking at the target frequency.
    (c) Transmission magnitude and phase for both directions are symmetric and nearly zero at 0.15 MHz, confirming complete energy dissipation for backward incidence and no loss for forward incidence. This confirms unidirectional perfect absorption (UPA).
    (d) Trajectories of the eigenvalues $\lambda_1$ and $\lambda_2$ of the scattering matrix in the complex plane, showing coalescence at an exceptional point (EP).
    (e–g) Frequency-dependent behavior of the eigenvalues’ imaginary parts (e), real parts (f), and magnitudes (g), confirming second-order EP characteristics near the same frequency.
    The response is achieved by independently tuning nine shunt circuit parameters, resistances ($R_1$, $R_2$, $R_3$), inductances ($L_1$, $L_2$, $L_3$), and strain-proportional voltage feedback gains ($V_{01}$, $V_{02}$, $V_{03}$) assigned to the three piezoelectric layers in the unit cell. }
    \label{fig:UPA_UZR_fmincon}
\end{figure}

An important question that arises is whether the reflection asymmetry is strong enough to achieve a reflection amplitude of zero in one direction and unity in the other. This scenario corresponds to the regime of \textit{unidirectional perfect absorption} (UPA), wherein a wave incident from one side is completely absorbed, while it is entirely retroreflected for incidence from the opposite side. To explore this regime, we formulate an optimization problem aimed at maximizing the reflection asymmetry, quantified as \( |r_f| - |r_b| \), by tuning the circuit parameters associated with the three shunted piezoelectric layers.

The objective is to determine the optimal combination of resistances, inductances, and voltage feedback gains that yields the maximum possible asymmetry, thereby enabling near-perfect reflection in one direction and zero reflection in the other. The optimization is carried out using MATLAB’s \texttt{fmincon} function, targeting the simultaneous realization of unidirectional zero reflection (UZR) and unidirectional perfect absorption (UPA) at the design frequency of 0.15\,MHz. 

The optimized circuit parameters are found to be:
\[
\{R_1, R_2, R_3\} = \{0,\, 29.28,\, 57.85\} \,\text{k}\Omega,\quad
\{L_1, L_2, L_3\} = \{0,\, 0.42,\, 1.20\}\,\text{H},\quad
\{V_0^1, V_0^2, V_0^3\} = \{7.66,\, 44.28,\, 12.06\} \,\text{MV}.
\]

It is important to note that this solution is not unique; alternative parameter combinations may also yield near-extreme asymmetry depending on the initial guess and convergence to local optima. The stochastic nature of the initialization process implies that different runs may lead to different but functionally equivalent solutions (see Methods section for more details). The results of the scattering analysis with these optimized circuit parameters applied to a single unit-cell scatterer are shown in Fig.~\ref{fig:UPA_UZR_fmincon}. The reflection amplitude and phase in Fig.~\ref{fig:UPA_UZR_fmincon}(a) demonstrate that the forward reflection is unity while the backward reflection is nearly zero, achieving UPA. The asymmetry metrics in Fig.~\ref{fig:UPA_UZR_fmincon}(b) further confirm this behavior, with a sharp peak in \( |r_f| - |r_b| \) and a phase contrast of approximately \( \pi \), consistent with EP conditions. Moreover, the transmission plots in Fig.~\ref{fig:UPA_UZR_fmincon}(c) indicate that transmission is suppressed at and beyond the EP frequency, while the backward reflection approaches unity in the high-frequency regime, realizing perfect retroreflection. The presence of an exceptional point is further validated by the eigenvalue evolution of the scattering matrix shown in Figs.~\ref{fig:UPA_UZR_fmincon}(d)–(g), where the eigenvalues coalesce and approach zero magnitude at the EP frequency, consistent with zero transmission, as predicted by Eq.~\eqref{eq:scattering_matrix}. This result demonstrates that extreme scattering asymmetry and critical-point behavior can be engineered entirely through circuit-based tuning, even in compact single-unit-cell structures.

\section*{Discussion}

Piezoelectric metamaterials with external shunt circuits provide a powerful avenue for controlling asymmetric wave phenomena using a fixed physical structure, unlike traditional elastic metamaterials that require geometric modifications. In this study, we demonstrate how resistive, inductive, and strain-proportional voltage feedback circuits embedded in a 1D layered piezoelectric composite can be used to precisely modulate asymmetric wave propagation. The composite structure supports Willis and electro-momentum coupling, as revealed through dynamic homogenization, which captures the nonlocal and non-Hermitian interactions emerging from broken inversion symmetry and circuit-induced loss and gain.

By parametrically tuning the shunt circuit components, we show that the degree of reflection asymmetry can be engineered to realize exotic wave phenomena such as unidirectional zero reflection (UZR) and unidirectional perfect absorption (UPA) at target frequencies. The voltage feedback, in particular, introduces an additional axis of control beyond the previously explored resistive and inductive loadings. This enhancement enables the realization of exceptional points (EPs) in the scattering matrix eigenvalue spectrum through purely circuit-based tuning, even within a single unit-cell configuration. The persistence of such critical-point behavior without periodicity underscores the localized nature of non-Hermitian scattering mechanisms.We employ a standard transfer matrix formalism to efficiently compute the scattering response and establish a homogenization framework in which the transfer matrix of a finite composite stack is matched with that of an equivalent medium governed by modified constitutive laws. The extracted effective parameters reveal nonzero Willis and electro-momentum coupling coefficients, highlighting the interplay between spatial asymmetry and non-Hermitian circuit control.

These findings provide a reconfigurable platform for nonreciprocal and asymmetric wave manipulation in compact structures, where critical phenomena like EPs can be accessed without requiring gain/loss balancing typical in parity-time symmetric systems. Although the demonstrated effects are frequency-specific, the circuit-based tuning approach enables real-time reconfigurability and could be extended to broadband and frequency-agile designs through adaptive or feedback-based control strategies. Future investigations will explore extensions to two- and three-dimensional structures, and experimental realization of programmable wave control in real time. These insights position circuit-coupled piezoelectric composites as a versatile and compact platform for implementing programmable non-Hermitian metamaterials, with potential applications in acoustic isolators, directional sensors, wave-based computing, ultrasonic imaging, and adaptive vibration control.

\section*{Methods}
\subsection*{Derivation of electro-mechanical elastic constant}
For a one-dimensional problem, the constitutive relations for each piezoelectric layer assuming scalar fields varying only along the \( z \)-direction can be written as:
\begin{equation}
\label{eq:const_law}
\begin{aligned}
    \sigma &= C \varepsilon - B E, \\
    D &= B \varepsilon + A E,
\end{aligned}
\end{equation}
where \( \sigma \) and \( \varepsilon \) are the longitudinal stress and strain, \( D \) and \( E \) are the electric displacement and electric field, and \( C \), \( A \), and \( B \) denote the elastic constant, dielectric permittivity, and piezoelectric coupling coefficient, respectively.

Assuming a transverse cross-sectional area \( A_p \), the current \( I \) through an external circuit and the voltage \( V \) across a layer are given by:
\begin{equation}
\label{eq:Elect_quat}
\begin{aligned}
    I &= \frac{\partial}{\partial t} \int_{A_p} D \, d\Omega_w, \\
    V &= \int_{l_p} E \, dz,
\end{aligned}
\end{equation}
where \( \Omega_w \) is the cross-section of the piezoelectric layer, and \( l_p \) is the layer thickness. Assuming a spatially uniform electric field and applying the Laplace transform with \( s = -i\omega \), Eq.~\eqref{eq:Elect_quat} becomes:
\begin{equation}
\label{eq:Elect_quat_Laplace}
\begin{aligned}
    I &= s D A_p, \\
    V &= E l_p.
\end{aligned}
\end{equation}Applying Kirchhoff’s voltage law to the shunt circuit, the voltage balance is:
\begin{equation}
\label{eq:Circuit_eq}
    V + I Z = V_0 \varepsilon,
\end{equation}
where \( Z = sL + R \) is the impedance of a series RL circuit with resistance \( R \) and inductance \( L \), and \( V_0 \) is the strain-proportional feedback gain. Combining Eqs.~\eqref{eq:const_law}, \eqref{eq:Elect_quat_Laplace}, and \eqref{eq:Circuit_eq}, we obtain the effective electromechanical constitutive relation:
\begin{equation}
\label{eq:ElectroMech_Const_Law}
    \sigma = \left[ C + \frac{s A_p B^2 Z}{s A_p A Z + l_p} - \frac{B V_0}{s A_p A Z + l_p} \right] \varepsilon = \check{C} \varepsilon,
\end{equation}
where \( \check{C} \) is the electromechanical elastic constant, which is tunable via the shunt impedance \( Z \) and the strain-proportional voltage \(V_0\). This effective constant \( \check{C} \) inherits the periodicity of the layered metamaterial and encapsulates the circuit-driven modulation of wave propagation behavior.

\subsection*{Scattering matrix calculations}
The transfer matrix for the \( i \)-th layer is given by:
\begin{equation}
\mathbf{T}_i =
\begin{bmatrix}
\cos(k_i l_p^i) & \dfrac{\sin(k_i l_p^i)}{\check{C}_i k_i} \\
-\check{C}_i k_i \sin(k_i l_p^i) & \cos(k_i l_p^i)
\end{bmatrix}, \quad
\begin{pmatrix}
u_i(z_i^R) \\
\sigma_i(z_i^R)
\end{pmatrix}
= \mathbf{T}_i
\begin{pmatrix}
u_i(z_i^L) \\
\sigma_i(z_i^L)
\end{pmatrix}
\label{eq:transfer_matrix}
\end{equation}

Here, \( k_i \) is the wavenumber in the \( i \)-th layer, and \( z_i^L \), \( z_i^R \) denote the left and right boundaries of the layer. The transfer matrix \( \mathbf{T}_i \) relates the state vector (displacement and stress) across each layer.

The displacement and stress fields within each layer are expressed as the sum of forward and backward traveling wave components:
\begin{align}
u_i(z, t) &= A_i e^{i(k_i z - \omega t)} + B_i e^{-i(k_i z + \omega t)}, \quad k_i = \omega \sqrt{\frac{\rho_i}{\check{C}_i}} \\
\sigma_i(z, t) &= i k_i \check{C}_i A_i e^{i(k_i z - \omega t)} - i k_i \check{C}_i B_i e^{-i(k_i z + \omega t)}
\end{align}
where \( A_i \) and \( B_i \) are the amplitudes of the rightward and leftward traveling waves, \( \rho_i \) is the density of the \( i \)-th layer, and \( \omega \) is the angular frequency.

The total transfer matrix for the finite composite, including aluminum layers on both sides, is given by:
\begin{align}
\mathbf{T}_f &= \mathbf{T}_0 \left( \mathbf{T}_1 \mathbf{T}_2 \mathbf{T}_3 \right)^n \mathbf{T}_0 \\
\mathbf{T}_b &= \mathbf{T}_0 \left( \mathbf{T}_3 \mathbf{T}_2 \mathbf{T}_1 \right)^n \mathbf{T}_0
\end{align}
corresponding to forward and backward wave incidence, respectively. Here, \( n \) denotes the number of unit cells in the composite, and each unit cell comprises three piezoelectric layers with different shunt circuit configurations. The reversal of the layer sequence in \( \mathbf{T}_b \) captures the asymmetry introduced by breaking spatial inversion symmetry.

The scattering coefficients are then obtained by enforcing continuity of displacement and stress at the aluminum–composite interfaces:
\begin{align}
\begin{pmatrix}
t_f \\ 0
\end{pmatrix}
&=
\begin{bmatrix}
1 & 1 \\
i C_0 k_0 & -i C_0 k_0
\end{bmatrix}^{-1}
\mathbf{T}_f
\begin{bmatrix}
1 & 1 \\
i C_0 k_0 & -i C_0 k_0
\end{bmatrix}
\begin{pmatrix}
1 \\ r_f
\end{pmatrix}
=
\mathbf{M}_f
\begin{pmatrix}
1 \\ r_f
\end{pmatrix}, \label{eq:Mf} \\
\begin{pmatrix}
t_b \\ 0
\end{pmatrix}
&=
\begin{bmatrix}
1 & 1 \\
i C_0 k_0 & -i C_0 k_0
\end{bmatrix}^{-1}
\mathbf{T}_b
\begin{bmatrix}
1 & 1 \\
i C_0 k_0 & -i C_0 k_0
\end{bmatrix}
\begin{pmatrix}
1 \\ r_b
\end{pmatrix}
=
\mathbf{M}_b
\begin{pmatrix}
1 \\ r_b
\end{pmatrix} \label{eq:Mb}
\end{align}

Here, \( C_0 \) and \( k_0 \) are the elastic modulus and wavenumber of the aluminum background. The matrices \( \mathbf{M}_f \) and \( \mathbf{M}_b \) encode the scattering response of the composite for forward and backward incidence.

\subsection*{Derivation of transfer matrix for dynamic homogenization}

The equation~\eqref{eq:State_Vector_EOM} defines a linear system of the form \( \vec{\Psi}'(z) = \mathbf{Q}(s) \vec{\Psi}(z) \), where \( \vec{\Psi}(z) \) is the state vector and \( \mathbf{Q}(s) \) is the system matrix dependent on material and circuit parameters. The corresponding transfer matrix \( \mathbf{T}(l_p) \) over a single piezoelectric layer of length \( l_p \) is given by:

\begin{equation}
\vec{\Psi}(z + l_p) = \mathbf{T}(l_p) \, \vec{\Psi}(z) = \exp\left[ \mathbf{Q}(s) \, l_p \right] \vec{\Psi}(z),
\end{equation}

where the exponential of the matrix \( \mathbf{Q}(s) \) governs the spatial evolution of the state vector due to wave propagation through the layer. The transfer matrix thus relates the electromechanical state at the two ends of the layer and captures the full coupled dynamic response.

To obtain the transfer matrix of a full unit cell, which may consist of multiple piezoelectric and elastic sublayers, the transfer matrices of the individual layers are sequentially multiplied in the order of wave propagation. For a three-layer unit cell with layer lengths \( l_p^1, l_p^2, l_p^3 \), the overall transfer matrix \( \mathbf{T}_{\mathrm{uc}} \) in the forward direction is given by:

\begin{equation}
\label{eq:unit_cell_transfer_matrix}
\mathbf{T}_{\mathrm{uc}} = \mathbf{T}_3(l_p^3) \, \mathbf{T}_2(l_p^2) \, \mathbf{T}_1(l_p^1),
\end{equation}

with total unit cell length \( l = l_p^1 + l_p^2 + l_p^3 \). Under the dynamic homogenization framework, this unit cell transfer matrix is equated to the exponential of an effective system matrix \( \tilde{\mathbf{Q}} \) acting over the entire unit cell:

\begin{equation}
\label{eq:unitcell_transfer_matrix}
\mathbf{T}_{\mathrm{uc}} = \exp\left[ \tilde{\mathbf{Q}} \, l \right],
\end{equation} where \( \tilde{\mathbf{Q}} \) encodes the effective dynamic properties of the homogenized unit cell. In the subsequent steps, this matrix is matched with that of the homogenized constitutive model by comparing scattering responses, enabling the extraction of effective parameters, including nonlocal terms such as Willis coupling and electro-momentum coupling coefficients. 

\subsection*{Computation of effective properties}

Starting with the effective constitutive laws defined in Eq.~\eqref{eq:Effective_Const_Law}, a governing equation similar in form to Eq.~\eqref{eq:State_Vector_EOM} can be derived for the homogenized medium. This formulation incorporates nonlocal coupling terms such as Willis and electro-momentum couplings, and relates the spatial derivatives of the averaged field variables to their values through a system matrix that captures the effective behavior. The resulting first-order differential system is given by:

\begin{equation}
\label{eq:Effective_EOM}
\frac{d}{dz}
\begin{pmatrix}
\langle u \rangle \\
\langle \phi \rangle \\
\langle \sigma \rangle \\
\langle D \rangle
\end{pmatrix}
=
\left[
\begin{array}{cccc}
\dfrac{-s \tilde{S}_D}{\tilde{C}_D} & 0 & \dfrac{1}{\tilde{C}_D} & \dfrac{\tilde{B}^\dagger}{\tilde{C}_D \tilde{A}} \\
\dfrac{s \tilde{W}_D}{\tilde{A}_D} & 0 & \dfrac{\tilde{B}}{\tilde{C}_D \tilde{A}} & -\dfrac{1}{\tilde{A}_D} \\
s^2 \left( \tilde{\rho} - \dfrac{\tilde{S}^\dagger \tilde{S}_D}{\tilde{C}_D} + \dfrac{\tilde{W}^\dagger \tilde{W}_D}{\tilde{A}_D} \right) & 0 &\dfrac{s \tilde{S}_D^\dagger}{\tilde{C}_D} & -\dfrac{s \tilde{W}_D^\dagger}{\tilde{A}_D}  \\
0 & 0 & 0 & 0
\end{array}
\right]
\begin{pmatrix}
\langle u \rangle \\
\langle \phi \rangle \\
\langle \sigma \rangle \\
\langle D \rangle
\end{pmatrix}
\end{equation}
where:

\begin{equation}
\label{eq:ModifiedEffectiveParams}
\begin{aligned}
\tilde{C}_D &= \tilde{C} + \frac{\tilde{B}^\dagger \tilde{B}}{\tilde{A}}, \quad 
\tilde{A}_D = \tilde{A} + \frac{\tilde{B}^\dagger \tilde{B}}{\tilde{C}}, \\[6pt]
\tilde{S}_D &= \tilde{S} + \frac{\tilde{B}^\dagger \tilde{W}}{\tilde{A}}, \quad
\tilde{S}_D^\dagger = \tilde{S}^\dagger + \frac{\tilde{W}^\dagger \tilde{B}}{\tilde{A}}, \\[6pt]
\tilde{W}_D &= \tilde{W} - \frac{\tilde{B} \tilde{S}}{\tilde{C}}, \quad
\tilde{W}_D^\dagger = \tilde{W}^\dagger - \frac{\tilde{S}^\dagger \tilde{B}^\dagger}{\tilde{C}}.
\end{aligned}
\end{equation}

The resulting parameters fully describe the dispersive, asymmetric, and nonlocal behavior of the homogenized piezoelectric composite with external circuitry. This system forms the foundation for constructing the transfer matrix of the homogenized domain, which captures the spatial evolution of the averaged electromechanical state vector. The transfer matrix can then be matched to that of the actual heterogeneous unit cell to extract the full set of effective parameters.

Specifically, by comparing the effective system matrix \( \tilde{\mathbf{Q}} \) in Eq.~\eqref{eq:unitcell_transfer_matrix} to the structure of the governing matrix in Eq.~\eqref{eq:Effective_EOM}, whose entries depend on the modified effective parameters defined in Eq.~\eqref{eq:ModifiedEffectiveParams}, the individual effective coefficients \( \tilde{C} \), \( \tilde{A} \), \( \tilde{B} \), \( \tilde{S} \), \( \tilde{W} \), and \( \tilde{\rho} \) can be systematically identified.

\subsection*{Circuit optimization}

Optimization was performed using MATLAB’s \texttt{fmincon} function with default solver settings and no additional options specified. The objective function,
\[
\text{obj} = |r_b| - |r_f|,
\]
was implemented as a custom function handle that returns the objective function value, where the design vector contains the nine circuit parameters, resistances ($R_1$, $R_2$, $R_3$), inductances ($L_1$, $L_2$, $L_3$), and voltage feedback gains ($V_0^1$, $V_0^2$, $V_0^3$) associated with the three shunted piezoelectric layers.

The initial guess for each optimization run was randomly generated using a scaled Gaussian distribution, i.e., \texttt{randn(9,1)$\times$100}, to span a broad range of possible values. To ensure physical feasibility, non-negativity constraints were enforced using a linear inequality of the form $-\mathbf{x} \leq 0$, implemented via the constraint matrix \texttt{-eye(9)} and the right-hand side vector \texttt{zeros(9,1)}. The optimization was repeated with multiple random initializations until the objective function approached a value near $-1$, corresponding to the regime of unidirectional perfect absorption and zero reflection. While the optimization yields a representative set of optimal parameters, the solution is not unique and may converge to different local optima based on the initial seed.

\bibliography{sample}
%\section*{Acknowledgements (not compulsory)}

%Acknowledgements should be brief, and should not include thanks to anonymous referees and editors, or effusive comments. Grant or contribution numbers may be acknowledged.

%\section*{Author contributions statement}

%Must include all authors, identified by initials, for example:
%A.A. conceived the experiment(s),  A.A. and B.A. conducted the experiment(s), C.A. and D.A. analysed the results.  All authors reviewed the manuscript. 

%\section*{Additional information}

%To include, in this order: \textbf{Accession codes} (where applicable); \textbf{Competing interests} (mandatory statement). 

%The corresponding author is responsible for submitting a \href{http://www.nature.com/srep/policies/index.html#competing}{competing interests statement} on behalf of all authors of the paper. This statement must be included in the submitted article file.

\end{document}